# Mobility Data Analysis and Applications
# A mid-year 2021 Survey

Abhishek Singh, Alok Mathur, Alka Asthana, Juliet Maina, Jade Nester,
Sai Sri Sathya, Santanu Bhattacharya, Vidya Phalke
[mprivacy.org](mprivacy.org)
Contact: [vidya.phalke@mprivacy.org](vidya.phalke@mprivacy.org)

## Abstract

In this work we review recent works analyzing mobility data and its application in understanding the epidemic dynamics for the COVID-19 pandemic and more. We also discuss privacy preserving solutions to analyze the mobility data in order to expand its reach towards a wider population.

## 1 Introduction

Various factors have emerged as indicators for understanding the spread of the COVID-19 and its impact on daily lives. One such important factor is mobility pattern. In this work, we survey and classify different ways in which mobility data has been utilized for various impactful initiatives, especially the current pandemic. We highlight different practical aspects of mobility data like regulations, curation, dissemination etc. We also discuss various privacy and ethics concerns associated with the mobility data collection and mining, and how they can be addressed.

### 1.1 Importance of mobility data

In 2020 over two-thirds of the world's population has mobile phones. This level of adoption coupled with the type of information available on a per user basis implies availability of tremendous data sets and insights. Location, applications, social networks, entertainment habits, data consumption and health are some of the key data sets that are rapidly becoming available. The global spread of COVID-19 pandemic early in 2020 catalyzed the usage of mobility data making it easier to collect, analyze and generate actionable insights.

### 1.2 Organization of this paper

The rest of the paper is structured as follows: First, we survey some of the common approaches to modeling mobility data. Then we present various research done on mobility analysis. Next, we discuss some of the applications and domains in which mobility data has been applied. Next, we discuss some of the privacy and ethics aspects of utilizing mobility data. Subsequently, we survey regulations across different regions that are relevant.

## 2 Data Modeling

**Defining mobility data:** In order to define mobility data, we look at different ways it can be obtained. One of the most common data collection methods is through users logging their GPS data and providing it to the crowd-sourced data collection efforts giving rise to spatio-temporal data. However, the GPS based data collection system requires a smartphone in majority of the cases and also a software capable of collecting the spatio-temporal data. Another mechanism for obtaining user level mobility data is through a top-down approach where the data is collected by service providers such as CDR data collected by telecommunication providers, foot traffic and points of interest data collected through sensors like WiFi usage, etc. The aforementioned data collection methods are not exhaustive and only covers some of the well known and widely used methods.

**Available datasets:** One of the widely used data providers is Safegraph [1],[2] that provides two major data sources. The first source is points of interest (POI) visits of various devices. The second source is the foot traffic data that gives a holistic picture of the mobility movement across different regions.

COVID-19 community mobility report [4] provides location specific mobility trends such as relative change in the foot traffic at a given place. This dataset is obtained by aggregation and anonymization of location data obtained by the Google maps users through the share location history feature.

Facebook data for good [34] provides various data streams relevant to mobility data such as population density maps, in-tile movement data, etc.

## 3 Mobility analysis using Telecom data

Call data record (CDR) has been used for different applications ranging from migration patterns, poverty spread, mobility patterns, impact of COVID-19 and etc. The [blog](blog) by Knipperken and Meyer describe two important insights and potential of CDRs - 1) They represent the population density and dynamics to a significant degree. 2) Analysis of the dynamics can result in insights about different segments of the population and the effect of intervention on those groups.

For definition, CDRs are data that are collected by mobile network operators (MNOs) for billing purposes, and are used internally for network optimization, new tower locations planning  etc.  A CDR is generated each time a call or SMS is made or received, and it includes an identifier of the SIM card, the timestamp of the call or SMS, and the location of the cell tower to which the call or SMS was passed through, typically one of the tower located closest to the user. As mobile phone penetration continues to increase globally, and especially in low- and middle-income countries, CDR analysis can be used to study movements of large numbers of people in an efficient way, free of interview bias.

During COVID-19, a pioneering country-wide study in Austria by Georg Heiler et. al [25] quantitatively assessed the effect of the lock-down for all regions of Austria, presenting insights of daily changes of human mobility by using near-real-time anonymized mobile phone data. Analyzing the mobility of population by quantifying mobile-phone traffic at specific point of interests (POIs), analyzing individual trajectories and investigating the cluster structure of the origin-destination graph, they found significant changes in the structure of mobility networks during the pre- and post- lockdown period. They also demonstrated the relevance of mobility data for epidemiological studies by revealing a significant correlation of the outflow from the town of Ischgl, an early COVID-19 hotspot in Austria, and the reported COVID-19 cases in Austria with an 8-day time lag.

An important validation when using CDR data is to ask whether it truly represents population, given that access to mobile phones itself may be skewed by people's ability to buy one, especially on poorer, vulnerable populations. Country-wide study in Gambia by Knippenberg and Meyer [26] tests this assumption. By correlating the known population density for each district against the density of unique phone users as defined by their International Mobile Equipment Identity (IMEI) prior to the confinement order (March 22, 2020) [26] they confirm a highly correlated relationship in both significance and magnitude, with population density as computed by WorldPop [27] and in the most recent population census. This validates the assumption that IMEI is a valid proxy for population density, and that tracking shifts in IMEI can therefore offer insights into short-term and long-term population movement dynamics.

Both these research indicates that mobile phone usage data permits the moment-by-moment quantification of mobility behavior for a whole country.

However, both these and many other studies also indicate the shortcoming of the current process in mobile data collection, aggregation, and analysis. First, the Telecom network and their monitoring tools are complex and require significant business knowledge. Second, the volume of the data tends to be enormous: typical MNOs in large countries can collect from tens of billions to hundreds of billions of records per day [28]. Third, the anonymization of the data that is required to be in compliance with several regional and country laws such as GDPR are complex, need to be employed at the source and are beyond the technical competencies of most researchers.

## 4 Applications and Domains

There are two broad categories of applications of mobility data - Predictive analysis and forecasting. While both categories are overlapping, there are significant differences in these two types of studies. Predictive analysis helps in explaining and quantifying past interventions and correlations in the mobility data. On the other hand, forecasting helps in estimating the future impact based on current and past trends in the mobility data. Beyond COVID-19, there are other domains severely impacting mobility patterns currently like climate change [21].

Effects on the disease spread [24] studies the impact of controlling human mobility as one of the interventions to control the disease spread. The authors perform epidemiological simulation by varying the mobility rate and highlight its correlation with infection rate and infection period. Forecasting [6]. Privacy preserving multi-operator contact tracing [3]. In [18] authors study the compliance for quarantine and lockdown by analyzing anonymised mobility patterns [14]

Change in mobility pattern due to the pandemic [16] highlights the relevance of performing data mining on relative change of mobility patterns instead of looking at absolute values. Their analysis reveals different patterns of mobility change across different age groups. In a separate study [10] researchers study the causal evidence for the impact of COVID-19 in mobility patterns for Sweden. In this study [17], authors build an anonymized location data to highlight daily changes in the points of interest location data. Socio-economic impact has been covered well in [5, 23] studying the correlation of economic condition with the mobile phone datasets.

## 5 Privacy and Ethics

There are various ethical and privacy concerns associated with sharing of mobility data. Some of these concerns affect individuals while others affect communities and business owners. A more detailed discussion of different ethical concerns associated with spatio-temporal data based contact tracing can be found here [22]

The privacy aspects of mobility data can be categorized as follows:

**Computational Privacy** based methods aim to deliver privacy by giving mathematical guarantees over either the algorithm or the input/output of the algorithm. There are different metrics to measure such privacy measures. One well known privacy definition is differential privacy [11] based on which several mechanisms have been proposed in the past. The underlying idea in differentially private mechanisms is to add sufficient noise to data such that it protects against presence of any particular individual in the dataset. The standard differential privacy based mechanisms require a trusted centralized authority which performs aggregation over data and then releases a noised version of data to untrusted entities. There are variants of differential privacy that operate under different threat models like local differential privacy [35] that do not require a trusted and centralized aggregator of the data sources. Typically local differential privacy has a relatively worse privacy-utility trade-off. Various recent works have proposed different ways of attaining privacy for spatio-temporal data such as location dependent privacy [36] which uses lipschitz privacy as the underlying privacy metric and adds noise proportional to the population density. Another notable work in this direction is geo-indistinguishability [37] which sets up a framework similar to lipschitz privacy.

While differential privacy may be useful for tasks where data release is the key requirement. There are cryptographic methods [38, 39, 40] that can be used for aggregation of mobility data under the same untrusted central server model but do not require any utility-privacy trade-off. Usually cryptographic techniques come at a higher cost of computation but have stronger security guarantees since they trade-off computation with privacy instead of utility (as done in differentially private mechanisms). Nevertheless, these techniques can not be used for releasing private data since the final release after decryption happens in plaintext. Another added advantage with differentially private mechanisms is that their results hold for computationally unbounded adversaries. However, all these methods come with strong assumptions like trusted servers, non-colluding parties, trusted enclaves etc. Therefore, the choice of the method to use depends on the threat model, systems architecture and what trade-offs stakeholders are willing to make for achieving privacy.

**Privacy by regulation** is an alternate model of enforcing privacy where the processing of information is performed by a trusted entity that follows the rules and regulations and does not perform any unauthorized computation. In the context of Covid-19, this has been used for many mobility data applications such as contact tracing [41]. Computational privacy is a safer and secure way of enforcing privacy but requires more careful and rigorous evaluation, hence both categories have their own trade-offs.

Typically, computational and mathematical tools for privacy are based on the principle of "cannot compromise privacy" in comparison to regulation enforced privacy where the underlying principle is "should not compromise privacy". Therefore both approaches provide their unique benefit along the spectrum of privacy-utility trade-off, trust in the system, etc.

## 6 Regions and Regulations

While the overall Telephony, Mobile and Internet standards across the world are the same or similar, the socio-economic and cultural norms are different. These differences show up in the way regulators provide directives and guidance. In this section of our survey we have highlighted some of the key regulatory or regulator influenced publications.

In Europe, from 2018, GDPR created a common framework and standard for data privacy in many situations. As such. The European Commission (EC) published a framework for developing a common approach for modelling and predicting the evolution of the coronavirus through anonymised and aggregated mobility data [30].

In the US, the regulatory picture is much more diverse and segmented across federal and state levels. The Congressional Research Service has a comprehensive document that describes 12 laws that directly deal with data privacy and are relevant as background material [31].

In India, there are two 2018 publications of relevance - the Policy Commission of India (NITI Aayog)'s National Strategy for AI in India [32], and the Telecom Regulatory Authority of India (TRAI)'s Recommendations on Privacy, Security and Ownership of the data in telecom sector [33]. Both these references go towards understanding the Indian landscape on the local regulations.

Finally, [12] provides a perspective from Nigeria. The paper describes an effective approach for digital contract tracing using mobile data which is compliant with Nigeria's National Data Protection Regulation (NDPR).

## 7 Conclusion and Future Work

In this work, we discuss different aspects of mobility data and its contribution in pandemic prevention technologies and other socio-economic factors. We also discuss various ethical concerns like privacy and briefly discuss some of the existing works addressing privacy concerns for geo-spatial data. We believe that standardized ways of ethically collecting such data streams could be extremely useful for various government and health institutions for addressing various challenges in a data driven manner.